\RequirePackage[l2tabu, orthodox]{nag}
\RequirePackage{snapshot}

\documentclass[9pt,onecolumn]{extarticle}

\makeatletter
\if@twocolumn
  \usepackage[dvips,letterpaper,top=0.5in, bottom=0.5in, left=0.75in, right=0.5in,includefoot,heightrounded]{geometry}
\else
  \usepackage[dvips,letterpaper,margin=1in,includefoot,heightrounded]{geometry}
\fi

\usepackage{srcltx}

\usepackage[russian,portuges,english]{babel}

\iflanguage{portuges}
    {\newcommand{\keywordname}{Palavras-chaves}}
    {\newcommand{\keywordname}{Keywords}}

\usepackage{amsmath}
\usepackage{amssymb,amsfonts}

\usepackage{abstract}

\usepackage{graphicx}
\usepackage[usenames,dvipsnames,svgnames,x11names]{xcolor}
\usepackage{subfigure}

\usepackage{booktabs}

\usepackage{setspace}
\usepackage{flushend}

\usepackage{cite}

\usepackage{hyperref}
\usepackage[normalem]{ulem}
\usepackage{bm}

\usepackage{enumerate}

\usepackage[noend]{algpseudocode}

\newcommand{\E}{\operatorname{E}}
\newcommand{\var}{\operatorname{Var}}
\newcommand{\diag}{\operatorname{diag}}

\usepackage{listings}

\usepackage[short,12hr]{datetime}
       \usepackage{fouriernc}
\makeatletter

\newcommand{\printtitle}{%
\makeatletter
\if@twocolumn

\twocolumn[%
  \maketitle
  \begin{onecolabstract}
    \myabstract
  \end{onecolabstract}
  \begin{center}
    \small
    \textbf{\keywordname}
    \\\medskip
    \mykeywords
  \end{center}
  \bigskip
]
\saythanks
\else
  \maketitle
  \begin{onecolabstract}
    \myabstract
  \end{onecolabstract}
  \begin{center}
    \small
    \textbf{\keywordname}
    \\\medskip
    \mykeywords
  \end{center}
  \bigskip
  \onehalfspacing
\fi
\makeatother
}

\author{%
B. G. Palm%
\thanks{Programa de P\'os-gradua\c{c}\~ao em Estat\'istica,
Universidade Federal Pernambuco, Brazil
and the Department
of Mathematics and
Natural Sciences,
Blekinge Institute of Technology,
Sweden
(E-mail: \protect\url{brunagpalm@gmail.com}).}
\and
F. M.~Bayer%
\thanks{Departamento de Estat\'istica
and LACESM,
Universidade Federal de Santa Maria, Brazil
(E-mail: \protect\url{bayer@ufsm.br}).
}
\and
R. J. Cintra%
\thanks{%
Signal Processing Group, UFPE, Brazil. (E-mail: \url{rjdsc@de.ufpe.br}).}
\and
M. I. Pettersson%
\thanks{Department of
Mathematics
and Natural Sciences,
Blekinge Institute of Technology,
Sweden
(E-mail: \protect\url{mats.pettersson@bth.se}).
}
\and
R. Machado%
\thanks{Department of
Telecommunications,
Aeronautics Institute
of Technology (ITA),
S\~ao Jos\'e dos Campos - SP,
Brazil
(E-mail: \protect\url{rmachado@ita.br}).
}
}

\title{%
Rayleigh Regression Model for
Ground Type Detection in SAR Imagery}

\newcommand{\myabstract}{%
This letter proposes a regression model for nonnegative signals.
The proposed
regression
estimates
the mean of Rayleigh
distributed signals
by a
structure
which includes
a set of regressors
and a link function.
For the proposed model,
we
present:
(i)~parameter estimation;
(ii)~large data record results;
and
(iii)~a detection technique.
In this letter,
we
present closed-form
expressions for the
score vector and
Fisher information
matrix.
The proposed
model
is
submitted to
extensive Monte Carlo
simulations
and to measured data.
The Monte Carlo
simulations
are
used
to
evaluate the performance of maximum likelihood estimators.
Also,
an
application
is
performed
comparing the detection
results of the
proposed model with
Gaussian-, Gamma-, and Weibull-based
regression models
in SAR images.
}

\newcommand{\mykeywords}{%
Detection,
Rayleigh distribution,
regression model,
reparameterized Rayleigh distribution,
SAR images.
}

\date{}

\begin{document}

\printtitle

\section{Introduction}

The classical linear regression model is commonly employed
to estimate
an
unknown
and
deterministic
parameter vector~$\bm{\beta}$
in the linear equation
$\mathbf{y} = \mathbf{H} \bm{\beta} + \mathbf{e}$.
The quantity~$\mathbf{y}$
is defined as the observed output signal,
$\mathbf{H}$ is a linear transformation,
and~$\mathbf{e}$
is a Gaussian noise vector~\cite{wiesel2008}.
However,
in situations
where the
observed output signal
is
asymmetric,
continuous,
and
nonnegative,
as in Rayleigh distributed signals,
inference methods
based on
the Gaussian assumption
can lead to
misleading results.
Indeed,
the Rayleigh distribution
is widely used
in
signal
and image
processing,
as
in~\cite{
oliver2004,
taricco2015,
zanetti2015,
gomes2016,
sumaiya2018, Lampropoulos1999}.

One important application
for the Rayleigh distribution
is
in the context of
synthetic aperture radar
(SAR) image
modeling,
where
this distribution
can be
employed for
characterizing
amplitude
values of image pixels~\cite{Kuruoglu2004,oliver2004,Jackson2009}.
A
common
problem
in SAR image processing
is
the
identification
and
classification
of
distinct
targets
or land uses
in images~\cite{Cintra2013,inglada2007}.
Usually,
these problems are treated
assuming homogeneity of the regions.
However,
the use of
regression models
adopting
suitable distributions
without assuming
homogeneity in the images
can
generate accurate results
for the above SAR-related challenges,
as presented by~\cite{wang2008}.
In this paper,
our goal is two-fold.
First,
we propose
a regression model
for
non-Gaussian
situations,
where the observed output signal
is
asymmetric
and
measured continuously
on the real positives values.
For the proposed
model,
we
introduce
parameter
estimation,
large data record results,
and
goodness-of-fit measures.
Second,
we
introduce
a change detector
for the amplitude values
of
non-Gaussian
SAR images.
Detection problems
are commonly
treated assuming
Gaussian distribution to the signals.
However,
SAR images
are usually non-Gaussian,
prompting the use of the Rayleigh distribution
to yield more accurate results for detection problems.
Thus,
the present letter
introduce a
detector
based on
the asymptotic properties
of the
proposed Rayleigh regression model estimators.

The letter is organized as follows.
In Section~\ref{s:model}, we introduce
the proposed model
and present
the score vector,
and the
goodness-of-fit measures.
Section~\ref{s:detec}
shows the
Fisher information matrix
and the proposed detector.
Section~\ref{s:vali}
presents
Monte Carlo
simulations
and
an application for SAR images.
Finally, the conclusion of this work
can be found in Section~\ref{s:conclusion}.

\section{Proposed Rayleigh Regression Model}
\label{s:model}

Let~$y$ be a random variable
with Rayleigh
distribution.
Its probability density function (pdf)
is given
by~\cite[p.~30]{Kay1998-2},~\cite{Jackson2009}:
\begin{align*}
p(y;\sigma)
=
\frac{y}{\sigma^2}
\exp\left(-\frac{y^2}{2 \sigma^2}\right)
,
\quad
y \geq 0
,
\end{align*}
where~$\sigma>0$ is the parameter.
The mean and the
variance of~$y$
are given by
\begin{align*}
\E(y) = \sigma \sqrt{\frac{\pi}{2}}
\quad
\text{and}
\quad
\var(y) = \sigma^2 \left(\frac{4-\pi}{2}\right)
.
\end{align*}
Although
the Rayleigh density
is commonly
governed by the
parameter~$\sigma$,
regression
models usually
characterize
the mean of the
response signal~\cite{McCullagh1989},
which
has a more
direct
interpretation
than~$\sigma$.
Thus,
we consider
a reparametrization of the Rayleigh distribution
in terms of the mean of the
response signal
and its regression structure.

\subsection{Reparametrization of the Rayleigh Distribution}

Considering the
parameterization~$\mu =\sigma \sqrt{\frac{\pi}{2}}$,
we have the following
pdf of the mean-based Rayleigh distribution:
\begin{align}
\label{e:rayleigh-rep}
f(y;\mu )
=
\frac{\pi y }{2 \mu ^2}
\exp\left(-\frac{\pi y^2}{4 \mu^2}\right)
,
\quad
y \geq 0
,
\end{align}
where~$\mu >0$ is the mean parameter.
The cumulative distribution function is given by
\begin{align*}%
F(y;\mu)= 1- \exp\left(-\frac{\pi y^2}{4 \mu^2}\right).
\end{align*}
The quantile function,
useful for
generating pseudo-random occurrences
in inversion method,
is given by
\begin{align*}
Q(u;\mu)
=
2 \mu \sqrt{ \frac{-\log(1-u)}{\pi} }
.
\end{align*}
The mean and variance
of~$y$
are given by
\begin{align*}
\E(y) = \mu
\quad
\text{and}
\quad
\var(y) = \mu^2 \left(\frac{4}{\pi}-1 \right)
.
\end{align*}

\subsection{Regression Model}

Let~$y[1], y[2], \ldots, y[N]$
be
$N$
independent random samples,
where each sample
follows
the
Rayleigh density in~\eqref{e:rayleigh-rep}
with mean~$\mu[n]$,
$n=1,2,\ldots,N$.
The proposed Rayleigh regression model is obtained
by considering
a linear predictor~$\eta[n]$
for the mean of~$y[n]$
furnished by
\begin{align}
\label{e:model}
\eta[n]
=
g(\mu[n]) = \sum_{i=1}^{r}  \beta_i  x_{i}[n]
,
\quad
n=1,2,\ldots, N,
\end{align}
where~$r<N$ is the
number
of covariates considered
in the model,
$\bm{\beta} = (\beta_1, \beta_2, \ldots, \beta_r)^{\top}$
is a
vector
of unknown
linear parameters,
$\mathbf{x}[n]=(x_{1}[n], x_{2}[n], \ldots, x_{r}[n])^\top$
is a
vector
of
deterministic
independent variables,
and~$g(\cdot)$ is a strictly monotonic and twice
differentiable link function where~$g:\mathbb{R}^+\!\rightarrow\mathbb{R} $.
If the intercept is considered,
then $x_1[n]=1$.
The link function~$g(\cdot)$
relates the linear predictors $\eta[n]$
to the expected value~$\mu[n]$ of data~$y[n]$.
When~$\mu[n]>0$,
a common choice of link function
is the log link
$\log(\mu[n])=\eta[n]$
with its inverse~$\mu[n]=\exp(\eta[n])$.

The proposed model is similar
to the generalized linear models
(GLM)~\cite{McCullagh1989},
except for the fact
that
the Rayleigh density
cannot be written
in the
canonical form of the exponential family of distributions.
A regression model
considering the
Rayleigh distribution
is also presented in~\cite{Aminzadeh1993}.
However,
the proposed model
is based on the
standard Rayleigh distribution
parametrization.
In addition,
in this letter,
the
maximum likelihood (ML)
method~\cite[Ch.~2]{Pawitan2001}
based on the reparametrized Rayleigh distribution
is considered to
obtain the
regression parameters
estimates,
as presented in the
next section.

\subsection{Likelihood Inference}

Parameter estimation of the Rayleigh
regression model can be performed
by
the maximum likelihood method~\cite[Ch.~2]{Pawitan2001}.
The ML
estimates are given by
\begin{align*}
\widehat{\bm{\beta}}
=
\arg
\max_{\bm{\beta}}
\ell(\bm{\beta})
,
\end{align*}
where~$\ell(\bm{\beta})$ is the log-likelihood
function
of the parameters
for the observed signal,
defined as
$
\ell(\bm{\beta})=\sum_{n=1}^{N} \ell[n](\mu[n])
.
$
The quantity~$\ell[n](\mu[n])$
is
the logarithm of~$f(y[n],\mu[n])$
given by
$
\ell[n](\mu[n])
=
\log\left(\frac{\pi}{2}\right) +
\log(y[n])
- \log(\mu[n]^2)-\frac{\pi y[n]^2}{4 \mu[n]^2},
$
where~$\mu[n]= g^{-1}\left(\sum_{i=1}^{r} x_{i}[n] \beta_i \right)$.

The score vector,
obtained by differentiating the log-likelihood function
with respect to each unknown parameters~$\beta_i$, is given by
$
U(\bm{\beta})=
\left(
\frac{\partial \ell(\bm{\beta})}{\partial \beta_1},
\frac{\partial \ell(\bm{\beta})}{\partial \beta_2},
\ldots,
\frac{\partial \ell(\bm{\beta})}{\partial \beta_r}
\right)^\top
.
$
Then, invoking the chain rule,
we have
\begin{align*}
\frac{\partial \ell(\bm{\beta})}{\partial \beta_i}=
\sum_{n=1}^{N}
\frac{d \ell[n](\mu[n])}{d \mu[n]}
\frac{d \mu[n]}{d \eta[n]}
\frac{\partial \eta[n]}{\partial \beta_i}
,
\end{align*}
where
\begin{align}
\frac{d \ell[n](\mu[n])}{d \mu[n]}
&= \frac{\pi y[n]^2}{2\mu[n]^3}-\frac{2}{\mu[n]}
,
\label{e:dldmu1}
\\
\frac{d \mu[n]}{ d \eta[n]}
\nonumber
&= \frac{1}{g'(\mu[n])},
\quad
\quad
\frac{\partial \eta[n]}{\partial \beta_i}
= x_{i}[n]
,
\end{align}
and~$g'(\cdot)$ is the first derivative of
the adopted link function~$g(\cdot)$.
In particular,
for the log link function,
$g(\mu[n])=\log(\mu[n])$,
we have~$
\frac{d \mu[n]}{d \eta[n]}
=\mu[n]$.

In matrix form,
the score vector can be written as
$
U(\bm{\beta})
=
\mathbf{X}^\top
\cdot
\mathbf{T}
\cdot
\mathbf{v}
,
$
where~$\mathbf{X}$ is an~$N \times r$
matrix whose~$n$th row
is~$\mathbf{x}[n]^\top$,
$
\mathbf{T}
=
\diag\left\{
\frac{1}{g'(\mu[1])},
\frac{1}{g'(\mu[2])},
\ldots,
\frac{1}{g'(\mu[N])}
\right\}
$
and~$\mathbf{v}
=
\left(\frac{\pi y[1]^2}{2 \mu[1]^3}-\frac{2}{\mu[1]},
\frac{\pi y[2]^2}{2 \mu[2]^3}-\frac{2}{\mu[2]},
\ldots,
\frac{\pi y[N]^2}{2 \mu[N]^3}-\frac{2}{\mu[N]} \right)^\top$.

The maximum likelihood estimators (MLEs) for the Rayleigh
regression parameters are obtained by solving the
following nonlinear system:
\begin{align}
\label{E:vescore}
U(\bm{\beta}) = \mathbf{0}
,
\end{align}
where~$\mathbf{0}$ is the~$r$-dimensional vector of zeros.
Solving~\eqref{E:vescore} requires the
use of nonlinear optimization algorithms.
We adopted
the
quasi-Newton Broyden-Fletcher-Goldfarb-Shanno
(BFGS) method~\cite{press}
for
the numerical computations.
We suggest
to use
as initial point estimate for
$\bm{\beta}$
the ordinary
least squares estimate
of $\bm{\beta}$,
obtained from a linear regression
of the transformed responses
$g(y[1],g(y[2]),\ldots,g([N])$
on~$\mathbf{X}$.

Based on the MLE of~$\bm{\beta}$,
it is possible to obtain a MLE
for~$\mu$,
considering the invariance principle
of the MLE~\cite[Ch.~2]{Pawitan2001},
as~$\widehat{\mu} = g^{-1} ( \mathbf{X} \bm{\widehat{\beta}})$.
\subsection{Goodness-of-fit Measures}

In this section,
diagnostic measures,
such as
the residual and
the coefficient of determination,
are
presented
to
evaluate the correct adjustment of the proposed model.
We considered the
quantile residual
as
$r[n]=\Phi^{-1}\left( F(y[n];\hat{\mu}[n])\right)$,
where~$\Phi^{-1}$ denotes the standard normal
quantile function.
The quantile residuals not only can
detect
poor fitting
in regression models but
its distribution is also approximately standard normal~\cite{Dunn1996}.

The
generalized coefficient of determination~\cite{nagelkerke1991},
which is
a global measure of the goodness-of-fit,
is given by
\begin{align*}
R^2  & = 1
-
\exp
\left(
- \frac{2}{N}
\left[
\ell ( \widehat{\bm{\beta}}) - \ell (\mathbf{0})
\right]
\right)
,
\end{align*}
where~$\ell (\mathbf{0})$
is the maximized log-likelihood of the
null model (without regressors)
and~$\ell (\widehat{\bm{\beta}})$
is the maximized log-likelihood of the fitted model.
Note that
$0 \leq R^2 \leq 1$
and
it indicates
the
proportion of the variability
of the observed output signal
that can be explained by the fitted model.
Higher values of~$R^2$
indicate more accurate predictions.

\section{Detection Theory}
\label{s:detec}

It is possible to interpret
a SAR image
as
a set of regions
composed of
possibly different types of
probability laws~\cite{Cintra2013}.
The problem of
correctly
distinguishing
between different
regions in one image
has been
studied
considering different
statistical approaches.
One approach to
achieve this goal
is the use of
the hypothesis test,
which allows for
the computation
of
differences
in
the mean of the amplitude
between
two separate regions
in
a given image~\cite{inglada2007,Cintra2013}.
In SAR image processing,
this technique
can also be considered
for
identification of
land cover type,
land cover
change detection or classification,
as shown
in~\cite{
mercier2008,hoekman2000}.

\subsection{Large Data Record Results}
\label{S:large}

Under some mild regularity
conditions~\cite[p.~167]{Kay1993},
the
MLEs
are consistent and
asymptotically~($N \rightarrow \infty$)
normally distributed.
Thus, for large data record,
\begin{align}
\label{e:normal}
\widehat{\bm{\beta}}
\sim \mathcal{N}_{r}
\left(
\bm{\beta}
,
(\mathbf{I}(\bm{\beta}))^{-1}
\right )
,
\end{align}
where~$\mathbf{I}(\bm{\beta})$
is the Fisher information matrix.
Their
asymptotic distribution
can be used to construct confidence intervals~\cite[Ch.~9]{Pawitan2001}
and
hypothesis tests~\cite[Ch.~9]{Pawitan2001}.

To obtain the Fisher information matrix
we need to calculate
the
expectation of the
negative value of the second-order partial derivatives of the
log-likelihood function~\cite[Ch.~8]{Pawitan2001}.
By applying the chain rule,
the second-order derivatives of the~$\ell(\bm{\beta})$
with respect to the~$\beta_i$, $i = 1,2,\ldots,r$,
are given by
\begin{align*}
\dfrac{\partial^2\ell(\bm{\beta})}{\partial\beta_i\partial\beta_p}
&= \sum_{n=1}^{N}\dfrac{d}{d\mu[n]}
\left( \dfrac{d \ell[n](\mu[n])}{d\mu[n]}
\dfrac{d\mu[n]}{d\eta[n]} \right)  \dfrac{d\mu[n]}{d\eta[n]}
\dfrac{\partial\eta[n]}{\partial\beta_p}   \dfrac{\partial\eta[n]}{\partial\beta_i} \\
& = \sum_{n=1}^{N}
\left(
\dfrac{\partial^2 \ell[n](\mu[n])}{\partial\mu[n]^2}
\dfrac{d \mu[n]}{d \eta[n]}
+ \dfrac{d \ell[n](\mu[n])}{d\mu[n]}
 \dfrac{\partial}{\partial \mu[n]}
\right. \\& \left.
 \times
  \dfrac{d\mu[n]}{d\eta[n]}
   \right)
 \dfrac{d\mu[n]}{d\eta[n]}
 \dfrac{\partial\eta[n]}{\partial\beta_p}   \dfrac{\partial\eta[n]}{\partial\beta_i}
,
\quad
i,p=1,2,\ldots,r
.
\end{align*}
Note that
taking expectation of \eqref{e:dldmu1},
we have that~$\E\left( d \ell[n](\mu[n])/d \mu[n] \right)=0$.
In addition,~$\dfrac{\partial\eta[n]}{\partial\beta_p}=x_{p}[n]$,
and~$\dfrac{\partial\eta[n]}{\partial\beta_i}=x_{i}[n]$.
Thus,
$
\E
\left[
\frac{\partial^2 \ell(\bm{\beta})}
{\partial \beta_i \partial \beta_p}
\right]
=
\sum_{n=1}^N
\left[
\E
\left(
\dfrac{d^2 \ell[n](\mu[n])}{d\mu[n]^2}
\right)
\left(
 \dfrac{d\mu[n]}{d\eta[n]}
\right)^2
x_{p}[n]x_{i}[n]
\right]
.
$
Now,
differentiating~\eqref{e:dldmu1},
we obtain
$
\frac{\partial^2 \ell[n](\mu[n])}{\partial \mu[n]^2}
= \frac{2}{\mu[n]^2} - \frac{3 \pi y[n]^2}{2 \mu[n]^4}.
$
Taking the expected value, we have
$
\E\left[\frac{d^2 \ell[n](\mu[n])}{d \mu[n]^2} \right]
= -\frac{4}{\mu[n]^2}.
$
Finally, we have
$
\E\left[ \frac{\partial^2 \ell(\bm{\beta})}
{\partial \beta_i \partial \beta_p} \right]=
 \sum_{n=1}^{N}
\left[
-\frac{4}{\mu[n]^2}
\left(
 \dfrac{d\mu[n]}{d\eta[n]}
\right)^2
x_{p}[n]x_{i}[n]
\right].
$
In matrix form, the Fisher information matrix is given by
$
\mathbf{I}(\bm{\beta})=
\mathbf{X}^\top
\cdot
\mathbf{W}
\cdot
\mathbf{X}
,
$
where
$\mathbf{W}
=
\diag\left\{
\frac{4}{\mu[1]^2}\left(\dfrac{d\mu[1]}{d\eta[1]}\right)^2,
\ldots,
\frac{4}{\mu[N]^2}\left(\dfrac{d\mu[N]}{d\eta[N]}\right)^2
\right\}$.

\subsection{Wald Test}

To test hypotheses over the regression parameters,
we partition the
parameter
vector
$\bm{\beta}=(\bm{\beta}_I^ \top , \bm{\beta}_M^\top)^\top$,
where~$\bm{\beta}_I$
is
the vector of parameters of interest with dimension~$\nu$
and
$\bm{\beta}_M$
is the nuisance parameter vector
with dimension~$r-\nu$.
The hypothesis of interest
is~$\mathcal{H}_0:\bm{\beta}_{I}=\bm{\beta}_{I0}$
versus~$\mathcal{H}_1:\bm{\beta}_{I} \neq \bm{\beta}_{I0}$.
Here,~$\bm{\beta}_{I0}$ is a fixed column vector of dimension~$\nu$.
The
Wald statistic can be written
as~\cite[p.~190]{Kay1998-2}:
\begin{align*}
T_W = (\widehat{\bm{\beta}}_{I1}-\bm{\beta}_{I0})^\top
\left( \left[ \mathbf{I} ^ {-1}
(\widehat{\bm{\beta}}_1) \right] _{\beta_I \beta _I} \right)^{-1}
(\widehat{\bm{\beta}}_{I1}-\bm{\beta}_{I0}),
\end{align*}
where~$\widehat{\bm{\beta}}_1 = (\widehat{\bm{\beta}}_{I1}^ \top ,
\widehat{\bm{\beta}}_{M1}^\top)^\top$
is the
MLE
under~$\mathcal{H}_1$
and
$
\left[ \mathbf{I}^{-1} (\widehat{\bm{\beta}}) \right]_{\beta_I \beta _I}
$
is a
partition
of~$\mathbf{I}(\widehat{\bm{\beta}})$
limited to the
estimates of interest.
From~\eqref{e:normal}
and based on the consistency of the MLE,
the~$T_W$
statistic
has an asymptotically
chi-squared distribution
with~$\nu$
degrees of freedom,
$\chi^2_\nu$.
The detection is performed by comparing the computed value
of~$T_W$
with
a threshold value~$\gamma$
obtained
from
the~$\chi^2_\nu$ distribution
and
the
desired
probability of false alarm~\cite{Kay1998-2}.

We assume that
the mean
of the
Rayleigh distributed signal
presents different values
depending on the ground type.
To illustrate,
consider
a region of forest in an image.
The detection
of this type of ground
can be obtained by
fitting the
following Rayleigh regression
model
$
g(\mu [n]) = \beta_1 + \beta_2 x_2[n]  + \sum_{i=3}^{r} \beta_i  x_i[n] ,
$
where
(i)~$\beta_1$ is the intercept;
(ii)~$x_2[n]$
is a binary covariate equal to one if the
region consists of forest and zero otherwise;
and
(iii)~$x_i[n]$, $i=3,4,\ldots,r$, are any other covariates that
can influence the mean of~$y$.
The detection problem
is
to distinguish between the
hypotheses:
\begin{align}
\label{e:detec}
\begin{cases}
\mathcal{H}_0 :
\mu [n] = g^{-1}
(\beta_1 + \sum_{i=3}^{r} \beta_i x_i[n]  ), \quad (\beta_2=0)
,
\\
\mathcal{H}_1 :
\mu [n] = g^{-1}
(\beta_1 + \beta_2 x_2[n]  + \sum_{i=3}^{r}  \beta_i x_i[n])
.
\end{cases}
\end{align}
To derive the detector,
we can use the
Wald test described above.
We
reject~$\mathcal{H}_0$
when~$T_W > \gamma$~\cite{Kay1998-2}.
In this situation,~$\beta_2 \neq 0$
and the
forest land use is detected.
This technique can be
considered to detect
any type of
ground in SAR images.

\section{Numerical Results}
\label{s:vali}

This section
presents
Monte Carlo simulations
and an
empirical investigation
in
ground type detection
in SAR images.
The Monte Carlo simulations
were used to evaluate the
MLE of the
Rayleigh regression parameters.
An application with real SAR data was considered
to demonstrate
the proposed detector.

\subsection{Analysis with Simulated Data}
\label{s:num}

The numerical results are based on the Rayleigh regression model
with the structure of the mean given by~\eqref{e:model}
considering the log
link function.
The parameters
were adopted as follows:
$\beta_1= 2$,~$\beta _2 = -1$, and~$\beta_3 = 1$
for Scenario~1,
and
$\beta_1= 0.5$ and~$\beta _2 = 0.15$
for Scenario~2.
The covariates
were generated
from the uniform
distribution~$(0,1)$
and considered constants
for all Monte Carlo replications.
In each replication
the inversion method
was considered to
generate~$y[n]$
assuming the Rayleigh distribution
with mean~$\mu[n]$.
The number of Monte Carlo replications
was set equal to~$10,000$ and the
signal
lengths
considered
were~$N \in \{
25, 250, 1,000
\}$.

We adopted the percentage relative bias (RB\%) and the means square
error (MSE) as figures of merit to numerically evaluate
the proposed point estimators.
Table~\ref{t:reg1}
presents the simulation results.
In general,
we notice that
the MLE
of the Rayleigh regression model
presented small
values
of percentage relative bias
and
mean square error.
As expected, increasing~$N$,
the percentage relative bias
and mean square error present lower values,
which matches the
consistence
of the MLE.
\begin{table}
\caption{
Results of the Monte Carlo simulation
of the point estimation for Scenarios~1 and~2
}
\label{t:reg1}
\centering
\begin{tabular}{c|ccc|cc}
\hline
&  \multicolumn{3}{|c|}{ Scenario 1} &   \multicolumn{2}{c}{ Scenario 2} \\
\hline
Measures 	&	$\widehat{\beta}_1$ 	&	$\widehat{\beta}_2$  	&
$\widehat{\beta}_3$ & $\widehat{\beta}_1$ 	&	$\widehat{\beta}_2$ 	\\
\hline
\multicolumn{6}{c}{ $N=25$ } \\
\hline
Mean & $1.9681$ & $-1.0030$ & $1.0040$ & $0.4810$ & $0.1472$ \\
RB(\%)   &  $1.5972$ &  $-0.3004$ & $-0.4045$ & $3.7913$ & $1.8467$ \\
MSE   & $0.0909$ &  $0.1564$ & $0.1533$ & $0.0470$ & $0.1421$ \\
\hline
 \multicolumn{6}{c}{ $N=250$ } \\
\hline
Mean &  $1.9971$ & $-1.0016$ & $1.0009$ &  $0.4984$ &  $0.1489$\\
RB(\%)   & $0.1450$ &  $-0.1600$ & $-0.0900$  & $0.3200$ &  $0.7333$ \\
MSE   &   $0.0073$ &  $0.0121$ &  $0.0126$ &  $0.0041$ &  $0.0125$ \\
\hline
\multicolumn{6}{c}{ $N=1,000$ } \\
\hline
Mean & $1.9993$ &  $-1.0001$ & $1.0002$ & $0.4995$ & $0.1502$\\
RB(\%)    & $0.0350$ &  $ -0.0100$ & $-0.0200$ &  $0.1000$ &  $-0.1333$\\
MSE   &  $0.0017$ & $0.0030$ & $0.0029$ &  $0.0010$ & $0.0030$ \\
\hline
\end{tabular}
\end{table}

\subsection{Analysis with Real Data}
\label{s:aplication}

The SAR image considered in this application
was taken by
CARABAS II~\cite{Lundberg2006},
a
Swedish UWB VHF SAR system.
The system
uses
HH polarization.
All information related to the data
can be found in~\cite{Ulander2005,Lundberg2006}
and the
images are
available in~\cite{data}.
The ground scene
of
the selected image
is
dominated by pine forest,
fences, power lines, military vehicles, and roads;
a lake
is also present~\cite{Lundberg2006}.

\begin{figure}
\centering
\includegraphics[scale=0.51]{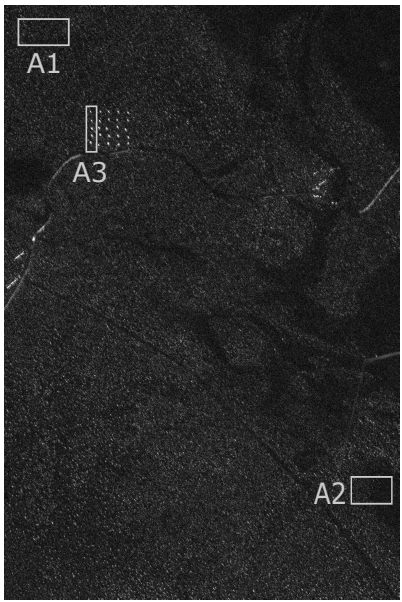}
\caption{CARABAS~II
single-look image
used in the regression models showing the regions tested.
Regions A1, A2, and A3
represent a forest, a lake, and an area
containing military vehicles,
respectively.}
\label{f:data}
\end{figure}

Figure~\ref{f:data}
shows
the three different regions
representing
forest,
lake,
and military vehicle
imagery;
referred to as
Regions~A1,~A2, and~A3,
respectively.
These regions were submitted
to the proposed modeling and detector.
The model is specified for the
mean of the response signal using
an intercept~($x_1[n] = 1$)
and two dummy variables~$(x_2[n] \,\, \text{and}\,\, x_3[n] )$
representing
each tested region,
as
$
g(\mu[n]) = \beta_1 +  \beta_2 x_2 [n]  + \beta_3 x_3 [n]
$.
The response signal is composed of the
amplitude values of the
pixels of the
Regions~A1,~A2, and~A3.
Variable~$x_2[n]$
is defined as one for
Region~A2 and zero for the rest.
The variable~$x_3[n]$ is
defined as
one for Region~A3 and
zero for the others.
Region~A1
is represented
when~$x_2[n]=0$
and~$x_3[n]=0$.

\begin{figure}
\centering
\subfigure[Residuals vs. index]
{\includegraphics[scale=0.72]{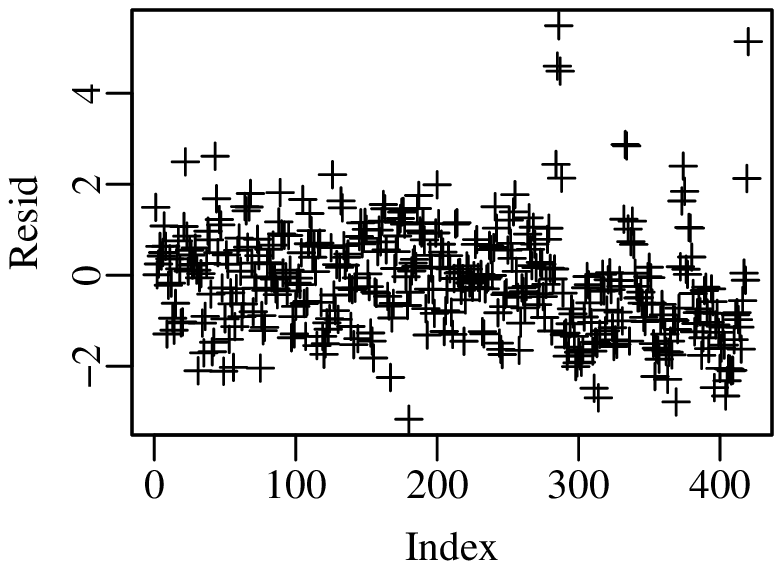}}
\subfigure[Histogram]
{\includegraphics[scale=0.72]{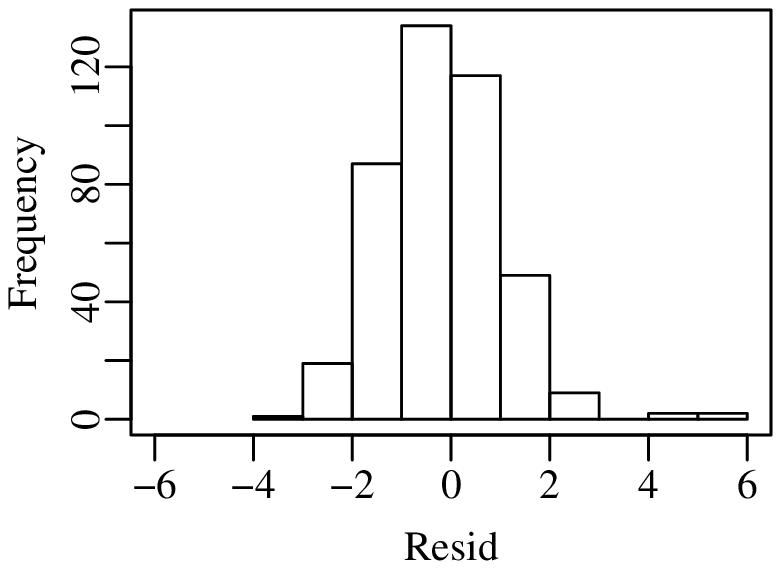}}
\caption{Residual charts for the Rayleigh regression model.}
\label{f:res}
\end{figure}

For comparison purposes,
we also
fitted
the standard Gaussian regression model,
the GLM with Gamma distribution,
and the Weibull regression model~\cite{zhang2016}
to the Regions~A1,~A2, and~A3.
Detection with Gaussian distribution
is widely discussed in literature
and the Gamma and Weibull distribution
are
also used
in SAR images, as in~\cite{vu2018,wang2006}.
The estimated parameters
for the considered models
are given in
Table~\ref{t:fit}.
In the Rayleigh regression model,
the mean response
presents a negative relationship
with $x_2[n]$
and positive relationship with $x_3[n]$.
Additionally,
we notice
that
the lake
and
the target regions
led
to
mean responses
which
are
$12.05\%$ lower
and
$194.00\%$ higher
than
the mean response from the forest region,
respectively.
The~$R^2$ values
of the fitted models
show that
the Rayleigh regression model
can explain~$70.96\%$
of the variation in~$y[n]$,
while the Gamma GLM, Gaussian,
and Weibull
regression models
can explain just~$30.09\%$,~$15.28\%$,
and~$0.3251\%$, respectively.
Figure~\ref{f:res}
presents the
residuals
of the Rayleigh regression model.
As expected, the residuals present values close to zero for
$98.81\%$
of
the observations
and
approximately
standard normal
distribution.

It is possible
to define a
detector
for this specific
regression model,
based on~\eqref{e:detec}.
The detection problem
in this image
is based
on computing the difference
in the behavior
among
the
tested regions.
With the~$p$-values
of the Wald test
presented in Table~\ref{t:fit},
we can verify that all variables
in the Rayleigh regression model
are significant for a probability of false alarm equal to~$0.05$.
Hence,
the
null
hypothesis
in~\eqref{e:detec}
can be rejected,
indicating a
correct detection of the land type.
In contrast,
the variable~$x_2[n]$
is not significant
for the
Gamma GLM,
Gaussian,
and
Weibull regression models,
i.e.,
the
Gaussian-,
Gamma-, and Weibull-based detections can not
distinguish the lake region from the other regions.
Thus,
the proposed Rayleigh regression model
can be used
for detecting
differences in SAR image regions
yielding
more accurate
results
when compared to
the
competing
regression models.

\begin{table}
\centering
\caption{Fitted regression models for
Regions~A1,~A2,
and~A3
}
\label{t:fit}
\begin{tabular}{lccc}
\bottomrule
\hline
& Estimate & Standard Error & Detection ($p$-value) \\
\bottomrule
\hline
\multicolumn{4}{c}{Rayleigh regression model}  \\
\bottomrule
\hline
$\widehat{\beta}_1$ &  $-2.0623 $ & $0.0445$ & $ < 0.001$ \\
$\widehat{\beta}_2$ & $-0.1280$ &   $0.0599$ &   $ 0.0325$  \\
$\widehat{\beta}_3$ &   $1.0784$ &  $ 0.0616$ & $< 0.001$ \\
\hline
\multicolumn{4}{c}{$R^2 = 0.7096$}  \\
\bottomrule
\hline
\multicolumn{4}{c}{Gaussian regression model}  \\
\bottomrule
\hline
$\widehat{\beta}_1$ & $0.12683$ &  $0.01646$  & $< 0.001$ \\
$\widehat{\beta}_2$ &   $-0.01201$ &  $0.02213$ & $0.588$  \\
$\widehat{\beta}_3$ &  $0.15948$ & $0.02277$   & $< 0.001$  \\
\hline
\multicolumn{4}{c}{$R^2 = 0.1528$}  \\
\bottomrule
\hline
\multicolumn{4}{c}{Gamma GLM}  \\
\bottomrule
\hline
$\widehat{\beta}_1$ &  $7.8844$ &  $ 0.5209$ &  $< 0.001$ \\
$\widehat{\beta}_2$ & $0.8248$ &   $0.7341$ &    $0.262 $ \\
$\widehat{\beta}_3$ & $-4.3917 $ &  $0.5657$ & $< 0.001$ \\
\hline
\multicolumn{4}{c}{ $R^2 = 0.3009$}  \\
\bottomrule
\hline
\multicolumn{4}{c}{ Weibull regression model}  \\
\bottomrule
\hline
$\widehat{\beta}_1$ & $-1.9939$ & $0.0583$ & $< 0.001$   \\
$\widehat{\beta}_2$ & $-0.1157$ & $0.0778$ & $0.1373$ \\
$\widehat{\beta}_3$ & $0.9583$ & $0.0815$ & $< 0.001$ \\
\hline
\multicolumn{4}{c}{$R^2 = 0.3251$}  \\
\bottomrule
\hline
\end{tabular}
\end{table}

\section{Conclusion}
\label{s:conclusion}

This letter introduced
a
new regression model for
nonnegative signals.
The proposed Rayleigh regression model
assumes that
the mean
of the
Rayleigh distributed signal
follows
a regression structure involving covariates,
unknown parameters,
and a link function.
An inference approach
for
the model parameters
is introduced
and diagnostic tools are discussed.
We also presented
Fisher information matrix,
asymptotic proprieties of the MLE,
and
a
detector
useful to
detect differences
in SAR image regions.
In the
Monte Carlo simulations,
the MLE of the
Rayleigh regression model showed
small values
of percentage relative bias
and mean square error.
An application of the Rayleigh regression model
to distinguish between different regions in a SAR image
was presented and discussed,
showing more accurate detection
results when compared with
the measurements from
Gaussian-,
Gamma-, and Weibull-based
regression models.

{\small
\singlespacing
\bibliographystyle{siam}
\bibliography{rayleigh}
}

\end{document}